\documentstyle{article}
\title{
\begin{flushright}
{\bf\normalsize   COLO-HEP-360}  \\
\end{flushright}
\vskip 10pt
\bf  Spin Glasses on Thin Graphs
}
\author{ {\it C.F. Baillie} \\
         Dept of Computer Science, University of Colorado\\
         Boulder, CO 80309, USA\\
         \\
         {\it W. Janke}\\ Institut fur Physik\\ Johannes
         Gutenberg Universitat\\
         D 55099 Mainz, Germany\\
         \\         
         \\
         {\it D.A. Johnston and P. Plech\' a\v{c}}\\
         Dept. of Mathematics\\
         Heriot-Watt University\\
         Riccarton\\
         Edinburgh, EH14 4AS, Scotland
}
\date {9 May 1995}         %if no \date --> current date
\begin{document}
  \maketitle
%-----------------------------------------------------------------------
                      {\Large
                      \begin{abstract}
%-----------------------------------------------------------------------
%
In a recent paper \cite{0} we found 
strong evidence from simulations that the Ising antiferromagnet on
``thin'' random graphs - Feynman diagrams - displayed
a mean-field spin glass transition. 
The intrinsic interest of considering such random graphs is that they
give mean field theory results
without long range interactions or
the drawbacks, arising from boundary problems, of 
the Bethe lattice. In this paper we 
reprise the saddle point calculations for the Ising and Potts ferromagnet,
antiferromagnet and spin glass
on Feynman diagrams. We use standard results
from bifurcation theory that enable us to treat
an arbitrary number of replicas and any
quenched bond distribution.  We
note the agreement between the ferromagnetic
and spin glass transition temperatures thus calculated and those
derived by analogy with the Bethe lattice or in 
previous replica calculations.

We then investigate numerically
spin glasses with a $\pm J$ bond distribution
for the Ising and $Q=3,4,10,50$ state Potts models,
paying particular attention to the independence
of the spin glass transition from the fraction
of positive and negative bonds in the Ising case
and the qualitative form of the overlap distribution $P(q)$ for all of
the models. The parallels with infinite range
spin glass models in both the analytical calculations and
simulations are pointed out. 
\\
Submitted to Nucl. Phys. B. [FS]  \\ \\
%
%-----------------------------------------------------------------------
                        \end{abstract} }
%-----------------------------------------------------------------------
%
  \thispagestyle{empty}
%
%***********************************************************************
%
  \newpage
%
%-----------------------------------------------------------------------
                  \pagenumbering{arabic}
%-----------------------------------------------------------------------

\section{Introduction and Analytical Calculations}
The analytical 
investigation of spin glasses on random graphs of various sorts has a long
and honourable history \cite{1,1a},
though there has been little in the 
way of numerical simulations \footnote{For some
recent simulations, see \cite{1b}.}. Random graphs with a fixed or fixed
average connectivity have a locally tree like structure, which means that
loops in the graph are predominantly large, so Bethe-lattice-like \cite{3} 
(ie mean field) critical behaviour is expected for spin models on such lattices. 
Given this, the analytical solution for a spin model
or, in particular, a spin glass on a Bethe lattice \cite{4,4a}
can be translated across to the appropriate 
fixed connectivity random lattice.
Alternatively, a replica calculation can be carried out directly
in some cases
for spin glasses on various sorts of random lattices.
For the case of Ising spins on a fixed connectivity random lattice we
arrive at 
the following prediction for the ferromagnetic transition temperature,
if one exists
\begin{equation}
\int_{-\infty}^{\infty} P ( J) \tanh ( \beta_{FM} J ) \; dJ = \frac{1}{z -1}
\label{e00}
\end{equation}
where $P(J)$ is the quenched probability distribution of bonds in the model, $z$ is the connectivity
and $\beta_{FM}$ is the (inverse) transition temperature.
A similar relation is predicted for the spin glass transition temperature
$\beta_{SG}$
\begin{equation}
\int_{-\infty}^{\infty} P ( J) \tanh^2 ( \beta_{SG} J ) \;  dJ = \frac{1}{z -1}.
\label{e01}
\end{equation}
From equs.(\ref{e00},\ref{e01}) 
we would expect to find a ferromagnetic transition when $P(J) = p \; \delta ( J - 1) +  (1 -p) \;\delta ( J + 1 )$
at 
\begin{equation}
 ( 2 p - 1 ) \tanh ( \beta_{FM}) = \frac{1}{ (z-1)}
\label{e1}
\end{equation}
or a spin glass transition at
\begin{equation}
\tanh^2 ( \beta_{SG} ) = \frac{1}{ (z-1)}.
\label{e2}
\end{equation}
depending on which critical temperature was lower.
The global order parameter that is necessary to describe the system in the spin glass
case \cite{1,1a} appears as the fourier transform of the local field distribution.

A rather different way of looking at the problem of spin models on random graphs was put forward in
\cite{5}, where it was observed that the requisite ensemble of random graphs
could be generated by considering
the Feynman diagram expansion for
the partition function of the model.
For an Ising ferromagnet with Hamiltonian
\begin{equation}
H = \beta \sum_{<ij>} \sigma_i \sigma_j,
\end{equation}
where the sum is over nearest neighbours on three-regular random graphs
(ie $\phi^3$ Feynman diagrams),
the partition function is given by
\begin{equation}
Z_n(\beta) \times N_n = {1 \over 2 \pi i} \oint { d \lambda \over
\lambda^{2n + 1}} \int {d \phi_+ d \phi_- \over 2 \pi \sqrt{\det K}}
\exp (- S )
\end{equation}
where $N_n$ is the number of undecorated graphs with $2n$ points,
\begin{equation}
N_n = \left( {1 \over 6} \right)^{2n} { ( 6 n - 1 ) !! \over ( 2 n ) !!
}
\end{equation}
$K$ is defined by
\begin{equation}
\begin{array}{cc} K_{ab} = & \left(\begin{array}{cc}
\sqrt{g} & { 1 \over \sqrt{g}} \\
{1 \over \sqrt{g}} & \sqrt{g}
\end{array} \right) \end{array}
\end{equation}
and the action itself is
\begin{equation}
S = {1 \over 2 } \sum_{a,b}  \phi_a  K^{-1}_{ab} \phi_b  -
{1 \over 3} (\phi_+^3 + \phi_-^3).
\label{e3}
\end{equation}
where the sum runs over $\pm$ indices.
The coupling in the above is $g = \exp ( 2 \beta J )$ 
where $J=1$ for the ferromagnet and 
the $\phi_+$ field can be thought of as representing ``up'' spins
with the $\phi_-$ field representing ``down'' spins.
An ensemble of $z$-regular random graphs would
simply require replacing the $\phi^3$ terms with $\phi^z$ and a fixed
average connectivity could also be implemented with the appropriate
choice of potential.

This approach was inspired by the considerable amount of work that has been done
in recent years on $N \times N$ matrix \footnote{$N$, the size
of the matrix 
is not to be confused with $n$, the number of vertices in the graph!}
versions of such integrals which generate
``fat'' or ribbon graphs graphs 
with sufficient structure to carry out a topological expansion \cite{6}
because of the matrix index structure. 
The natural interpretation of such fat graphs as the duals 
of triangulations, quadrangulations etc. of surfaces has led to much interesting work in string
theory and particle physics \cite{7}. 
The partition function here is a poor, ``thin''
(no indices, so no ribbons), scalar cousin of these,
lacking the structure to give a surface interpretation to the graph. Such scalar
integrals have been used in the past 
to extract the large $n$ behaviour of various
field theories \cite{8} again essentially as
a means of generating the appropriate Feynman diagrams, so a lot is known about handling their quirks. 

For the Ising ferromagnet on three-regular ($\phi^3$) graphs, 
solving the saddle point equations at large $n$
shows that the critical behaviour appears as an exchange of dominant saddle point solutions
to the saddle point equations
\begin{eqnarray}
\phi_+ &=&  \sqrt{g} \phi_+^2 + {1 \over \sqrt{g}} \phi_-^2 \nonumber \\ 
\phi_- &=&  \sqrt{g} \phi_-^2 + {1 \over \sqrt{g}} \phi_+^2 
\end{eqnarray}
at $g = \exp (2 \beta_{FM}) = 3$.
The high and low temperature solutions respectively are
\begin{eqnarray}
\phi_+,\phi_- &=&  {  \sqrt{g} \over  g + 1   
         }  \nonumber \\
\phi_+,\phi_- &=& { \sqrt{g} \over 2 ( g - 1)} \left( 1 \pm   \sqrt{ { g  - 3 \over g + 1} } \right)
\end{eqnarray}
which give a low temperature magnetized phase that can be detected
by a non-zero magnetization order parameter, which is defined in terms of the
fields $\phi_+,\phi_-$ as
\begin{equation}
M = { \phi_+^3 - \phi_-^3 \over \phi_+^3 + \phi_-^3}.
\end{equation}
The critical exponents for the transition can also be calculated
in this formalism and, as expected, are mean field. In general
a mean field transition appears at 
\begin{equation}
\exp ( 2 \beta_{FM}) = z / ( z - 2)
\end{equation}
on $\phi^z$ graphs, 
which is exactly the value predicted by
equ(\ref{e1}) in the standard approaches.

To consider a non-trivial bond distribution it suffices
to make the substitution \cite{5}
\begin{equation}
K_{ab} \rightarrow \int_{-\infty}^{\infty} K_{ab} \; P ( J ) \; d J.
\label{e4}
\end{equation}
in the saddle point equations.
If we take the distribution $P(J) =  p \; \delta ( J - 1) + (1 - p) \; \delta ( J + 1 ) $,
this has the effect of replacing $g$ by
\begin{equation}
{ p g + 1 - p \over p + ( 1 - p ) g }
\end{equation}
On $\phi^z$ graphs we find that this change in the coupling shifts the ferromagnetic
transition point to
\begin{equation}
\exp ( 2 \beta_{FM} ) = {  (2 p - 1) ( z - 1) + 1 \over  ( 2 p - 1) ( z - 1 ) - 1  }
\label{ex}
\end{equation}
which is again identical to the value found from equ.(\ref{e1}).

This is rather surprising as
a direct substitution of the 
weighted propagator coming from equ.(\ref{e4})
into the saddle point equations 
might be expected to 
correspond an {\it annealed} distribution of bonds
because we are calculating with a finite
number (one) of replicas. 
The calculations leading to equ.(\ref{e1}) and equ.(\ref{e2}), however,
have taken quenched bond distributions by calculating
with $k$ replicas and then taking the limit $k \rightarrow 0$.

\section{The Saddle Point Equations for Spin Glasses}

There are various possibilities for addressing
spin glass order in the Feynman diagram approach.
In \cite{5} the entropy per spin was calculated
for the Ising {\it anti}-ferromagnet on $\phi^3$ graphs and
it was found to become negative for sufficiently negative
$\beta$, which is often indicative of a spin glass transition.
Similarly it was found that the factorized solution 
(which exists in the ferromagnetic and anti-ferromagnetic cases) 
broke down for
higher moments $Z^k$ of the partition function, which is again
indicative of a spin glass transition.
The temperature at which this happened appeared to be
converging to a finite value  as the moments increased,
unlike the random energy model where a similar calculation
gives a temperature that diverges as $\sqrt{k}$ \cite{9}.

We now look at the calculation of $Z^2$
in a little more detail.
We can put two Ising models on each Feynman diagram 
by taking four fields $\phi_{++}, \phi_{+-}, \phi_{-+}, \phi_{--}$,
where the double subscript now covers both replicas,
and enlarging the inverse propagator to
\begin{equation}
\begin{array}{cccc} K_{ab} = & \left(\begin{array}{cccc}
g & 1 & 1 & {1 \over g} \\
1 & g & {1 \over g} & 1 \\
1 & {1 \over g} & g & 1 \\
{1 \over g} & 1 & 1 & g 
\end{array} \right) \; . \end{array}
\end{equation}
If we solve explicitly the saddle point equations 
for the two
replica system with $P(J) = p \delta (J -1) + (1 - p) \delta ( J + 1 )$
we find a $p$-independent solution appearing at
$g=5.8284..$ and a mirror solution at the corresponding
antiferromagnetic value given by the inverse of this, $g=0.1716...$. 
The second solution, $g=0.1716...$,
was also observed 
in \cite{5} as the point
where the factorized solution broke
down for the antiferromagnet
with $k=2$. Remarkably, as we
are looking at a finite number (two) of replicas here, 
$g=5.8284,g=0.1716$ are precisely the values given by  
eqn.(\ref{e2}) for $\beta_{SG}$, where the limit $k \rightarrow 0$
has been taken to obtain the spin glass transition
temperature. Thus, just as for the ferromagnetic transition, the
quenched $k \rightarrow 0$ results are appearing already at $k \ne 0$.
Using numerical routines to investigate the structure of the saddle
point equations for more Ising replicas reveals that the solutions
at $\beta_{FM}$ and $\beta_{SG}$ still appear at higher $k$ as
bifurcations from the symmetric high temperature solution.

This is not quite the whole story. As already noted in \cite{5}
for $k=3,4...$ we hit a different sort of critical behaviour
which takes us off the symmetric solution branch
to a {\it replica} symmetric solution
before reaching $\beta_{SG}$. At these points
a first order (jump in $q$) transition occurs. 
This behaviour is
identical to that of the Ising replica magnet \cite{9a},
which is effectively the SK model at a finite number of replicas
for a suitable choice of parameters. In this the $k=2$ transition is
second order and the transition temperature coincides with the $k=0$
value, whereas the $k \ge 3$ transitions are first order and occur
at higher temperatures.

We can attempt to analyse
the saddle point equations for {\it any} number $k$ of Ising replicas in the case
of purely ferromagnetic or antiferromagnetic couplings using 
bifurcation theory methods because the tensor product structure of 
the  propagator $K_{ab}$ is preserved in these cases
\footnote{This is just a more formal statement of the
observation in \cite{5} that a solution given by the product
of $k=1$ solutions existed in these cases.}
which facilitates the calculation. 
We will not, however, see first order transitions such as those discussed
above because the bifurcation structure
only tells us about continuous
transitions from the symmetric high temperature solution.

If we denote the
$2 \times 2$  propagator in equ.(7) as $K(g)$ for brevity, the saddle point equations
for the $k$ replica case may now be written schematically as
\begin{equation}
{\partial S_k \over \partial \vec \phi} \simeq \vec \phi -  \otimes^k K(g)  \vec \phi^2 = 0
\end{equation}
where $\vec \phi, \vec \phi^2$ are now $2^k$ dimensional vectors, the tensor
product is taken over $k$ copies of $K$
and the action for $k$ replicas is 
denoted by $S_k$. 
We have abused the notation slightly in equ.(18) as the left hand side is
really $\otimes^k K(g) \partial S_k / \partial \vec \phi$ because
we have multiplied by
$\otimes^k K(g)$ to get rid of the inverse factor coming from
the quadratic term in $S_k$, which simplifies the analysis
of the saddle point equations.
We will continue the abuse in the discussion below.

For the antiferromagnetic case
$g$ is simply replaced by $1/g$ throughout.
We restrict ourselves to investigating
the symmetric branch (and hence ignore 
the first order solutions that appear for $k>2$).
A symmetric, high temperature solution can be constructed
by taking products of the $k=1$ symmetric solution in equ.(10) $(t,t)$
where 
\begin{equation}
t = {\sqrt{g}  \over g + 1 }.
\end{equation}
Standard theory \cite{bif} then shows that a bifurcation
from the symmetric high temperature solution (ie phase transition) is expected
when the Hessian $\det \left( \partial^2 S_k / \partial \vec \phi^2 \right)$ is equal to zero. More explicitly
\begin{equation}
\det \left({\partial^2 S_k \over \partial \vec \phi^2} \right)= \det ( 2 t^k K_k -1)
\end{equation}
where we have denoted the tensor product of $k$ copies of $K$ by $K_k$. If we absorb
the factors of $t$ into $K$ and denote $\tilde K = ( \sqrt{g} + 1 / \sqrt{g})^{-1} \; K$
we thus have
\begin{equation} 
\det \left({\partial^2 S_k \over \partial \vec \phi^2} \right) = \det ( 2 \tilde K_k -1) = 0
\end{equation}
as our bifurcation equation. 
The extension to $z$-regular ($\phi^z$) graphs is trivial. For these the 
$k=1$ high temperature solution is ($\tilde t, \tilde t$) where
\begin{equation}
\tilde t = \left( {\sqrt{g} \over g + 1} \right)^{1/(z -2)}
\end{equation}
and the Hessian becomes
\begin{equation}
\det \left({\partial^2 S_k \over \partial \vec \phi^2} \right) = \det ( (z-1) (\tilde t^{z-2})^k K_k -1)
\end{equation}
so the net effect is simply to replace $2$ by $z-1$ in the
bifurcation equation. 

The matrix $\tilde K$ has two eigenvalues $1$
and $(g -1)/(g+1) = \tanh (\beta)$. 
From the algebraic structure of $\tilde K_k$
we can readily calculate the eigenvalues $\lambda_m = \tanh (\beta)^m$, 
$m = 0,1,2,\dots k$
where each eigenvalue has multiplicity ${{k} \choose {m}}$  and express
(taking $z=3$)
\begin{equation}
\det \left({\partial^2 S_k \over \partial \vec \phi^2} \right) = \prod_{m=0}^{k} (2 \tanh (\beta)^m - 1)^{{k} \choose {m}} = 0.
\end{equation}
The roots of the bifurcation equation are thus
\begin{equation}
\tanh (\beta) = 2^{-1/m} \; \; m = 1,2, \ldots k
\end{equation}
reproducing equ.(3) for $p=1$ and $z=3$.
In terms of $\beta =  (1 / 2) \log (g)$ we can write this as
\begin{equation}
\beta_m = { 1 \over 2}
\log \left( { 2^{1/m} + 1 \over 2^{1/m} -1} \right)  \; \; m = 1,2 \ldots k
\end{equation}
where we have retained the label $m$ on $\beta$. The above analysis holds for the ferromagnet
where $g = \exp(2 \beta) \ge 1$. In the case of a pure antiferromagnet $g = \exp(2 \tilde \beta) \le 1$
and $\tanh(\beta)$ is negative, so we can only find roots of the bifurcation equation for $m$ even.
\begin{equation}
\tilde \beta_m = { 1 \over 2}
\log \left( { 2^{1/m} - 1 \over 2^{1/m} + 1} \right)  \; \; m = 2, 4 \ldots k
\end{equation}
We have thus recovered both $\beta_{FM}=\beta_1$ ($g=3$) 
for the ferromagnet and
$\beta_{SG}=\tilde \beta_2$ ($g=0.1716$) for the antiferromagnet
in this approach. 

A generic feature of the above solutions
is immediately obvious - once a bifurcation appears at some $k$ it
is always present for a greater number of replicas. This suggests
a possible resolution of the puzzle of why the quenched
transition temperatures are appearing already at finite $k$:
The appropriate transition is encountered at the
same temperature for {\it all} $k$, so the analytical
continuation $k \rightarrow 0$ is trivial. In the spin glass case
the first order transitions for $k \ge 3$ are the fly in the
ointment, as these are encountered before the putative continuous
transition at $\beta_{SG}$. Looking explicitly at the
replica symmetric solution for $k=3$ \cite{5} shows that the
the replica symmetric banch
appears in a first order transition at $g \simeq 0.19 \ldots$
(in the antiferromagnet) before the
symmetric branch takes over again below this.

It is rather amusing that we have encountered a ``c=1 
(2 Ising model) barrier''
in this context. Calculations
are still possible here for $k>2$, but there is
certainly a change of behaviour. In the case of
spin models on annealed ensembles of fat graphs
analytical calculations have struggled to get
beyond $c=1$, whereas simulations have so far
failed to see much difference between $c<1$
and $c>1$.

Although the tensor product structure of $K_k$ is lost when
$p \ne 0,1$ and the saddle point equation becomes
\begin{equation}
{\partial S_k \over \partial \vec \phi} = \vec \phi - \left( p \otimes^k K(g)  + ( 1 -p ) 
\otimes^k K(1/g) \right) \vec \phi^2 = 0
\end{equation}
the matrices $K(g),K(1/g)$ differ only in the {\it sign} of the eigenvalue
$\tanh (\beta)$, so it is still possible to derive an explicit expression for the Hessian,
which in this case is
\begin{equation}
\det \left({\partial^2 S_k \over \partial \vec \phi^2} \right) = \prod_{m=0}^{k} \left[
2 \tanh (\beta)^m 
\left( p + (-1)^m ( 1 - p)\right) - 1 \right]^{{k} \choose {m}} .
\end{equation}
We thus reproduce the values of $\beta_{M}$ and $\beta_{SG}$ calculated from equs.(3,4) for general $p$ as the first two bifurcation points. On $\phi^z$ graphs we replace $2 \rightarrow z-1$, which is still in accordance with equs.(3,4).
As all
even bifurcation points are independent of $p$, we predict the $p$
independence of the spin glass transition point, as in the more standard
approaches. 

Equs.(3,4) are not restricted to distributions of the form $P(J) = p \delta (J -1) 
+ (1-p) \delta (J + 1)$, but we can extend the approach here to accommodate this
by making the observation that $K(g_1)$ will commute with another $K(g_2)$
for any $g_1,g_2$.
Thinking of a given distribution (a gaussian, say) as a weighted sum of delta
functions over $K$'s with different values of $g$, we can still diagonalize all
of the terms in the sum simultaneously and arrive at an expression of the form
\begin{equation}
\det \left({\partial^2 S_k \over \partial \vec \phi^2} \right) = \prod_{m=0}^{k} \left[  2 \int P(J) \tanh (\beta J)^m dJ - 1 \right]^{{k} \choose {m}} = 0.
\end{equation}
for the bifurcation equation.

Although the continuous transition temperatures $\beta_m$ are independent
of the number of replicas, the multiplicity of solutions
(ie low temperature phases) that bifurcate at a given point are 
not. 
There are $k ( k - 1) / 2$ solutions bifurcating at the spin glass
transition temperature $\beta_2$, so just as in the infinite
range model there is a change in the nature of the solutions to
the saddle point equation when $k < 1$ and the number of solutions
is formally negative. The classification of the spin glass
solutions in the current case
is thus clearly closely related to the parametrization
of the matrix $Q_{ab}$ that appears in the saddle point equations
of the infinite range model \cite{sk}.

It is also possible to consider Potts spins on a Bethe lattice \cite{4a,4b}.
We take Hamiltonian in this case to be
\begin{equation}
H =  2 \beta \sum_{<ij>} \delta_{\sigma_i, \sigma_j}
\end{equation}
where the spins $\sigma_i$ can now take on $Q$ values
\footnote{We have adopted
a different normalization for the Potts Hamiltonian from
\cite{4a}, taking a factor of 2 rather than $Q$ in front,
for consistency with \cite{0}}. 
The critical temperatures for ferromagnetic and spin glass ordering 
calculated in \cite{4a} give, 
in the style of equs.(\ref{e00},\ref{e01}),
\begin{equation}
\int_{-\infty}^{\infty} P(J) \left( { \exp(2 \beta_{FM} J) - 1 
\over \exp(2 \beta_{FM} J) + Q - 1 } \right) dJ = { 1 \over z - 1}
\label{e6}
\end{equation} 
and
\begin{equation} 
\int_{-\infty}^{\infty}P(J) \left( { \exp(2 \beta_{SG} J) - 1 
\over \exp(2 \beta_{SG} J) + Q - 1 } \right)^2 dJ = { 1 \over z - 1}   
\label{e7}
\end{equation}
for $Q$ state Potts models. If we consider the ferromagnet, $P(J)= \delta ( J - 1)$,
we find that $\exp( 2 \beta_{FM}) = ( z + Q - 2 ) / ( z - 2 )$.
There is, however, a parallel here with the Ising spin glass
calculations. It was noted in \cite{4b} that this
temperature is actually a spinodal point, for $Q>2$ the Potts
ferromagnet undergoes a first order transition before reaching this
point.
Arguing, as in the Ising case, that a random graph looks locally like
the Bethe lattice we would expect the various critical temperatures
and spinodal points still to apply on $\phi^z$ graphs.

It is also possible to solve the saddle point equations explicitly
to avoid relying on this analogy.
For the 3-state Potts model with action
\begin{equation}
S = { 1 \over 2 } ( \phi_a^2 + \phi_b^2 + \phi_c^2 ) - c ( \phi_a \phi_b + \phi_a \phi_c + \phi_b \phi_c)
-{1 \over 3} ( \phi_a^3 + \phi_b^3 + \phi_c^3 ),
\label{potts3}
\end{equation}
this
gives high and low temperature solutions\footnote{We gave these, and those for the 4-state model,
incorrectly in \cite{0}.}
\begin{eqnarray}
\phi_{a,b,c} &=& 1 - 2 c \; ; \nonumber \\
\nonumber \\
\phi_{a,b} &=& { 1 + \sqrt{1 - 4 c - 4 c^2} \over 2} \; ,\nonumber \\
\phi_c &=& { 1 + 2 c - \sqrt{1 - 4 c - 4 c^2} \over 2}
\label{potts3sol}
\end{eqnarray}
where $c = 1/ ( g + 1)$. This 
gives a changeover in behaviour at $c = 1/5$, ie $g=4$, which agrees
with the value from equ.(\ref{e6})
and thus picks out the spinodal point. 
Similarly, solving the 4-state Potts model equations
gives the solutions
\begin{eqnarray}
\phi_{a,b,c,d} &=& 1 - 3 c \; ; \nonumber \\
\nonumber \\
\phi_{a,b,c} &=& { 1 + 3c - \sqrt{1 - 6 c - 3 c^2} \over 2} \; ,\nonumber \\
\phi_d &=& { 1 - c + \sqrt{1 - 6 c - 3 c^2} \over 2}
\end{eqnarray}
where $c = 1/ (g+2)$ here. The changeover here is at $g=5$, again
apparently at the spinodal point.
It would be interesting to carry out a simulation to check
the nature of the transition in these ferromagnetic models.

Analysis of the Potts spin glass saddle point equations follows a similar tack to those
of the Ising model. 
The caveats about missing potential 
first order transitions  at finite $k$ still, of course, apply.
If we take the three state Potts model as an example $K$ is 
replaced by the $3 \times 3$ matrix $L(g)$
\footnote{This definition of $L$, which is the simplest from the
point of view of the saddle point equations, involves a rescaling
with respect to equ.(\ref{potts3}) which accounts for
the different between $t$ and the high temperature solution 
in equ.(\ref{potts3sol}).}
\begin{equation}
\begin{array}{cc} L_{ab} = & \left(\begin{array}{ccc}
1 & { 1 \over g} & {1 \over g}\\
{1 \over g} & 1 & {1 \over g} \\
{1 \over g} & {1 \over g} & 1  
\end{array} \right) \end{array}
\end{equation} 
and the high temperature solution on $\phi^3$ graphs is ($t,t,t$) where
\begin{equation}
t = { g \over g + 2 }.
\end{equation}
The eigenvalues of $\tilde L(g) = t L(g)$ are $1,\lambda,\lambda$, 
where $\lambda = ( g -1 ) / ( g + 2)$.
It is still true that $\tilde L(g_1), \tilde L(g_2)$ will commute for any $g_1,g_2$ so we can put this
to good use again to derive the Hessian for general $P(J)$
\begin{equation}
\det \left({\partial^2 S_k \over \partial \vec \phi^2} \right) = \prod_{m=0}^{k} \left[  2 \int P(J) \left(\exp(2 \beta J) - 1 \over \exp(2 \beta J) +2
\right)^m dJ - 1 \right]^{{k \choose m} 2^m}.
\end{equation}
The multiplicity of solutions has changed because of the possibility of taking
one of two $\lambda$ from each replica. The results for $\beta_{FM}$ and $\beta_{SG}$ from
the Bethe lattice are thus faithfully reproduced as the first two bifurcation points.

Generalizing to the $Q$ state Potts model on
$\phi^z$ graphs, we find
\begin{equation}
t = \left({ g \over g + Q -1 }\right)^{1/(z-2)}
\end{equation}
and the eigenvalues of the $Q \times Q$ matrix
$\tilde L$ will be $1$ and $Q-1$ $\lambda$'s, where
$\lambda = ( g -1) / ( g + Q -1)$. This gives the Hessian
\begin{equation}
\det \left({\partial^2 S_k \over \partial \vec \phi^2} \right) = \prod_{m=0}^{k} \left[  ( z -1)
 \int P(J) \left(\exp(2 \beta J) - 1 \over
\exp(2 \beta J) + Q -1
\right)^m dJ - 1 \right]^{{k \choose m} (Q-1)^m}
\end{equation}
which the reader will be pleased to hear is the last elaboration of the bifurcation
equation that we consider. The $Q=2$ case reproduces the Ising results derived earlier. 

Given the accessibility of mean-field results on random
graphs, especially as one is not forced to take infinite range interactions
or deal with the boundary difficulties of the Bethe lattice,
the lack of simulational effort is rather surprising.
As the new
viewpoint
offered by regarding the random graphs as Feynman diagrams 
offers the possibility of attacking various 
questions, such as
the nature of replica symmetry breaking on random graphs, from a different angle, we thought it important 
to continue the simulations of \cite{0}
for 
the Ising and $Q$ state Potts spin glasses on $\phi^3$ graphs 
to check carefully the agreement with the
known results for quantities such as the transition temperatures
and critical exponents and to gain a thorough understanding
of the numerics before going on to investigate less
well understood aspects. 
Our numerical methods are largely similar to those of
\cite{0}, so we describe them only briefly in the next section and concentrate
on the results of the simulations.

\section{Simulations: Generalities and Ising Model}

In all of the simulations we report here we generated
around 100 different Feynman diagrams at each of the sizes
we simulated, 100, 250, 500 and 1000 vertices. The quenched
bond distribution on each graph was $P(J) = p \; \delta ( J -1)
+ ( 1 - p) \; \delta ( J + 1 )$, with $p$
taken to be $0.5$ (the "$\pm J$"
distribution) unless specified otherwise. 
Various
massively parallel processors were used to run the simulations
of the different graphs simultaneously and perform the
quenched averaging over the graphs in situ. Again, 
in the interests of simplicity and reliability (and as in
\cite{0}) we used only a metropolis algorithm
with simulated annealing rather than some of the more advanced
algorithms developed recently. At each $\beta$ value simulated
we carried out 500,000 simulated annealing sweeps, followed
by 500,000 production sweeps with a measurement every tenth
sweep. Each sweep consisted of a complete metropolis
update of the lattice. Our strategy for extracting the 
critical exponents will be to use Binder's cumulant for
the overlap, defined below, to extract an estimate for $\beta_{SG}$
and the combination $\nu d$ and then to look at the finite size scaling
of other quantities to extract further exponents.

In the Ising model the spin glass transition temperature
is found in simulations by putting two Ising replicas
on each graph with spins $\sigma_i, \tau_i$
to measure the overlap, which is the order parameter for the 
spin glass transition
\begin{equation}
q = {1 \over n} \sum_i \sigma_i \tau_i.
\end{equation}
The Binder's cumulant for the overlap is then
defined in an analogous fashion to the
Binder's cumulant for the magnetization in a ferromagnetic transition
\begin{equation}
U_{sg} =  { [ \langle q^4 \rangle ] \over [ \langle q^2
\rangle ]^2 }
\label{cumq}
\end{equation}
where $\langle \  \ \rangle$ denote thermal averages and
$[ \ \ ]$ disorder averages. The plots of $U_{sg}$
for differently sized graphs are expected to cross at $\beta_{SG}$.
In the course of all the simulations the 
overlap distribution 
\begin{equation}
P_n(q) = \left[ \langle \delta ( q - 1/ n \sum \sigma_i \tau_i) \rangle \right]
\end{equation}
was also histogrammed. A non-trivial 
$P(q)$ is a strong, but not infallible, signal for a spin glass phase.
It is perhaps
worth remarking that the order of averages in eqn.(\ref{cumq})
is not absolutely obvious {\it a priori} - one might have considered
$[ \ < q^4> / <q^2>^2 ]$, for instance. Regarding $<q^n>$ as
moments of the distribution $P(q)$, the choice in the eqn.(\ref{cumq})
would seem most appropriate. The alternative
overall average has been considered
in quantum spin glass simulations \cite{10} on the heuristic
grounds that it gives
better scaling behaviour. We have carried out the scaling analysis
of our simulations
described below with both definitions for the Ising spin glass
and found essentially identical results.

We expect the overlap to have the finite size scaling form
$q \simeq n^{-\beta / \nu d}$ at any spin glass transition
and the spin glass susceptibility, defined as
\begin{equation}
\chi_{sg} = { 1 \over n} \sum_{ij} [ \langle \sigma_i \sigma_j
\rangle^2 ]
\label{csg1}
\end{equation}
or, alternatively,
\begin{equation}
\chi_{sg} = n \int q^2 P_n ( q ) dq.
\label{csg2}
\end{equation}
to diverge as $\chi_{sg} \simeq n^{\gamma / \nu d}$. Other
critical exponents
such as that for the 
specific heat
$C \simeq B + C_0 n^{\alpha / \nu d}$ are defined in the standard manner
and relations such as $\alpha = 2 - \nu d$ still
hold good. 

The mean field critical exponents are
shown in the table below for an Ising ($Q=2$) spin glass
\vspace{.1in}
\centerline{Table 1:  Spin Glass exponents for Ising model}
\begin{center}
\begin{tabular}{|c|c|c|c|c|c|c|c|c|c|c|c|c|c|c|c|c|c|} \hline
$\alpha$  & $\beta$  & $\gamma$ & $\delta$ & $\nu d$ & $\eta $\\[.05in]
\hline
$-1$  & $1$ & $1$ & $2$ & $3$   & $0$\\[.05in]
\hline
\end{tabular}
\end{center}
\vspace{.1in}
We have written $\nu d$ in the table rather than $\nu$
as we cannot disentangle $\nu$ and $d$ in the infinite
dimensional random graph case at hand.

We now commence the analysis of the results of the simulations proper
for the Ising model. The cumulant for the overlap is plotted
using both ways of carrying out
the averages in Fig.1 and Fig.2, from which it is clear that the resulting
crossover points are, as claimed, very similar.
Fitting the crossover point from either figure gives
$\beta_{SG}=0.88(1)$, although the overall average that gives
Fig.2 is slightly noisier. 
This value is in good agreement with the 
theoretical predictions. An 
additional consequence of equ.(\ref{e2}) is that $\beta_{SG}$
should be independent of $p$ for $P(J) = p \delta (J - 1 )
+ ( 1 - p) \delta ( J + 1 )$. 
In \cite{0} we considered
the Ising antiferromagnet, which has $p=0$,
and found $\beta_{SG} = -0.94(2)$ 
by direct observation of cumulant crossing, but we also found
$\beta_{SG} = -0.88(2)$ from extrapolation of the specific heat peak. 
These results are thus only marginally compatible with the
$p$ independence. As a further check we simulated a
complete set of graphs with $p=0.3$ and found that the 
cumulants were identical within the errors to the $p=0.5$ values,
this giving the same $\beta_{SG}$. The slight discrepancy at 
$p=0$ may well be due to the poorer statistics we have 
for these runs.

Further analysis of the cumulants allows us to
extract $\nu d$ by looking at the scaling of the maximum
slope of the cumulant with the number of vertices $n$.
\begin{equation}
max \left( { d U \over d \beta } \right) \simeq n^{1 \over \nu d}
\end{equation}
We find that $\nu d = 2.9(1)$ from the data in Fig.1 and
$\nu d = 3.0(2)$ from that in Fig.2. The simulations in \cite{0}
found the compatible value of $\nu d = 2.8(2)$ for $p=0$
and all are in agreement with the mean-field value of $\alpha= -1$
for the specific heat critical exponent,
deduced from $\alpha = 2 - \nu d$. A cusp in the 
specific heat $C$, 
rather than a divergence,
is indicated by this negative value of $\alpha$ in
the finite size scaling relation
\begin{equation}
C \simeq B + C_0 n^{\alpha / \nu d}
\label{esh}
\end{equation}
The cusp can clearly be seen in Fig.3. A direct fit to the 
scaling of the specific heat as in eqn.(\ref{esh})
is not particularly instructive, as the extra adjustable constant $B$
allows for a very good fit to the value $\alpha / \nu d = - 1/3$,
but it is at least consistent with the value emerging from the cumulant analysis.

Turning now to the expected scaling of the overlap, we find 
respectable agreement with the finite size scaling relation 
\begin{equation}
q \simeq n^{-\beta / \nu d}
\end{equation}
with $\beta / \nu d$ calculated to be $0.30(1)$
by extrapolating to the pseudocritical value of $\beta_{SG}=0.88$.
Having considered the first moment of $P(q)$ with $q$, we now move on to
what is effectively the second moment of the $P(q)$ with 
$\chi_{sg}$. In Fig.4
we see the expected divergence, and a fit to the finite size scaling
relation
\begin{equation}
\chi_{sg} \simeq n^{\gamma / \nu d}
\end{equation}
gives $\gamma / \nu d = 0.34(2)$, in agreement with the 
mean field value of $\gamma =1$.
No divergence is expected in the {\it linear} susceptibility
\begin{equation}
\chi_M = {d M \over d H}
\end{equation}
and, as is clear from Fig.5, none is seen.
We have not measured the response of the model to an external
field, nor attempted to fit the correlation functions at the critical
temperature, so we have no estimates for $\delta$ and $\eta$.
However, it is clear from the above results that
the model is giving mean-field like behaviour for the spin glass
transition that is observed.

The results for the specific heat and the susceptibilities
reported here for the $\pm J$ spin glass are essentially
identical to those in \cite{0} for the Ising antiferromagnet,
which can be taken as a further confirmation of the
$p$ independence of the spin glass transition. 
Looking directly
at the histograms of $P(q)$ themselves in the spin glass phase for $p=0$ (the antiferromagnet)
and $p=0.3,0.5$ (the models simulated here) does show some 
differences as can be seen in Fig.6.
However, the determination of the critical point depends on the scaling
of the moments of $P(q)$ around the transition point 
at $\beta=0.8814...$ and it is here we expect to find $p$ independence.
This is indeed what we see at $\beta=0.9$ in Fig.7 - the histograms
are essentially identical for $p=0.3,0.5$. There is a slight
discrepancy still for $p=0$, which is a reflection of the less convincing scaling
analysis in \cite{0} that determined the critical point in the case of the
pure antiferromagnet. We have shown only the $P(q)$ for graphs with 500 points in
the figures for clarity, the results for other graph sizes are identical.

\section{Simulations: Potts Models}

For $Q$-state Potts models the overlap may be
defined as
\begin{equation}
q = { 1 \over n} \sum_{i=1}^n ( Q  \delta_{\sigma_i , \tau_i} - 1 )
\end{equation}
which is arranged to be zero in the uncorrelated case. 
The cumulants and $P(q)$ may then be defined in an
analogous fashion to the Ising model.
There
is, however, no longer the $q \rightarrow -q$ symmetry in $P(q)$
that is present in the Ising spin glass
for $Q \ge 3$ state Potts models. 
In all of the simulations reported here we took
$P(J) = 1/2 ( \delta ( J - 1 ) +  \delta ( J + 1 ))$.
We simulated lattices of size $100,250,500$ and $1000$ for
the $Q=3,4$ state models, but only the three smaller lattice
sizes for $Q=10,50$. The statistics and annealing schedules
were identical to the Ising model simulations.
We are principally interested in observing the qualitative features 
of the various Potts models, so our analysis concentrates less on 
serious finite size scaling to extract the exponents than the 
Ising results.

Mean field theory (in the infinite range model) suggests
that $Q \ge 3$ Potts glasses are rather different in behaviour
from the Ising spin glass as there is no longer
continuous replica symmetry breaking. 
For $Q=3,4$ there are two
consecutive transition temperatures $\beta_{SG}$
and $\beta_{SG2}$ to different glass
phases and for $Q>4$ the first has
a discontinuity in the overlap $q$. 
For $\beta_{SG2} > \beta > \beta_{SG}$ $P(q)$ consists
of a delta function at $q=0$ and another
delta function at finite $q$ with no continuous features
whereas for $\beta > \beta_{SG2}$ the delta function at finite $q$
splits in two with a continuous
distribution between.
The values 
of the transition temperature given
by equ.(33,41) are for $\beta_{SG}$. 
Some features of the solution differ 
on the Bethe lattice \cite{4a}, in particular an Ising-like solution
appears to exist for coordination number three with a $\pm J$
bond distribution, and the $\phi^3$ random graph model might be
expected to behave in this fashion too.

If we take the 3 and 4 state Potts models first, we can look at the cumulant crossing
in a similar manner to the Ising model to attempt to pinpoint 
the phase transition.
It is 
possible to extract crossing points at $\beta=1.5(1)$ for $Q=3$ and $\beta=2.0(1)$
for $Q=4$, which are in agreement with the values
calculated from equ.(\ref{e7}) shown below in Table.2. 
Our data points 
for the $Q=10,50$ state models are rather sparser
but the cumulant crossing appears to become
more clear cut with increasing $Q$, which offsets this.
\vspace{.1in}
\begin{center}
\begin{tabular}{|c|c|c|c|c|c|} \hline
$Q$& 2 & 3  &  4 & 10  & 50  \\[.05in]
\hline
$\beta_{SG}$& 0.8814 & 1.5824  & 2.1226 & 3.6914 & 6.1943 \\[.05in]
\hline
$measured$& 0.88(1) & 1.5(1)  & 2.0(1) & 3.6(4) & 6.1(1) \\[.05in]
\hline
\end{tabular}
\end{center}
\vspace{.1in}
\centerline{Table 2:  $\beta_{SG}$ for various $Q$ state Potts models on $\phi^3$ graphs.}
\vspace{.1in}
The quoted errors in the table are the most conservative
choice - the largest difference between the crossing points
for the various lattice sizes - apart from $Q=3$ where the
$N=100$ data was dropped, as it failed to cross with the
other lattice sizes.

We next look at the qualitative
features of the low temperature phase in the various $P(q)$
to see if the mean field theory expectations are borne out.
With our definitions, 
the histograms of $P(q)$ range from $-1$ to $Q-1$ for a
$Q$ state Potts model with the origin at zero and roughly
half of the
probability density in the range $0 \ldots (Q-1)$, so we have
plotted only $0 \ldots (Q-1)$, normalized
to the range $0 \dots 1$ for clarity in all our figures
\footnote{It is also possible to extract a positive scalar
order parameter for the spin glass phase by going to a
simplex representation for the Potts spins and using the 
radial component of the overlap matrix, but we stick with
the canonical definition here.}
In all of the models the high-temperature $P(q)$ look
like skewed gaussians centred on the origin,
whose sharpness increases with graph size
indicating delta-function-like behaviour in the continuum limit. 
In Fig.8 we have plotted $P(q)$ for the three state Potts
model for $\beta=2.2$ ($>\beta_{SG}$). This is clearly
different from the Ising model: there is a strong
peak at the origin that is increasing with graph size
as well as smaller bump at around $q=0.7$ which 
is also increasing with graph size and which might tentatively
be identified with another peak in the continuum limit.
In \cite{0} simulations of the Potts antiferromagnet, where
no spin glass phase is expected, saw no signs of the structure at $q=0.7$,
with $P(q)$ remaining resolutely centred on the origin. 
Although the very large autocorrelation times that are encountered
for $\beta>\beta_{SG}$ make the interpretation of the low
temperature results a dangerous business we can press on deeper
into the low temperature region to look at $P(q)$ there and see
if there is any evidence for the second transition. $P(q)$ is
plotted at $\beta=3.5$ for the same three lattice sizes in Fig.9,
where we can see that there does, indeed, appear to be a 
second peak developing away from the origin at $q=0.2$
as well as the bump at $q=0.7$.

The qualitative features of the above results
are preserved in the $Q=4,10,50$ state models that we
also simulated: the histogram of $P(q)$ broadens from a sharp peak
at the origin at around $\beta_{SG}$ and develops a 
secondary bump or shoulder. There is some evidence of
further structure developing for even larger $\beta$.
From the numerical evidence it would thus appear that the $Q=3$
Potts spin glass with a $\pm J$ distribution of bonds lies
{\it above} the critical $Q$ value that separates Ising and Potts
spin glass behaviour, whereas the Bethe lattice calculation in
\cite{4a} supports the opposite conclusion, although
on the relatively small lattices simulated here
it is difficult to distinguish between features
that will be sharp in the continuum limit and continuous parts
of $P(q)$.
A more detailed analysis of the saddle point equations may
shed some light on this question analytically. 
For completeness in Fig.10
we show $P(q)$ for the $Q=4$
model at $\beta=4.4 (>\beta_{SG})$ to demonstrate the similarity with the
$Q=3$ results.

The fits to the maxima of the Binder's cumulant to extract $\nu d$ are
rather poor for all the higher state Potts models, but the specific heat
curves for $Q=3,4$ are 
similar in form to the Ising model, as can be seen for
the $Q=4$ Potts model in Fig.11, indicating the continued
presence of a cusp rather than a divergence. We therefore do not have
a reliable value of $\nu d$ to feed into the finite size scaling
relations,
although the cusp suggests that $\nu d \simeq 3$, at least for
$Q=3,4$.
Nonetheless, we can still fit to find 
the combinations $\beta / \nu d$
and $\gamma / \nu d$ at the estimated critical points
for the various models. This gives $\beta / \nu d = 0.28(2),0.31(5)$
for the $Q=3,4$ state models respectively. The mean field theory suggests
that $q$ becomes discontinuous at $\beta_{SG}$ for $Q>4$. With the smaller
lattices that we have for $Q=10,50$ this effect, if it exists,
is masked by finite size rounding but attempting to fit $\beta / \nu d$
does give larger values and much poorer fits than for $Q=3,4$
such as $0.45(5)$ for $Q=10$.
This could be construed as providing some evidence for a discontinuity.
The data allowed fits to $\gamma / \nu d$ only for the $Q=3,4$
models, where it gave $0.36(2), 0.40(7)$ respectively. 

In summary, all the Potts $P(q)$ are clearly qualitatively
different from the Ising model,
even though the two exponents $\beta / \nu d, \; \gamma / \nu d$
we fitted for $Q=3,4$
are roughly similar. There is not sufficient
data in the simulations reported here to reliably distinguish
a discontinuous transition for $Q>4$, though there are
certainly indications (the fits to $\beta / \nu d$) that this is the case.
In all the cases, however, the values extracted for the spin glass
transition temperature by analysis of Binder's cumulant
are in agreement with the predictions.

\section{Conclusions}

The novel analytical approach offered by using techniques
borrowed from matrix models in the manner of \cite{5}
to look at spin models on random graphs allows one to
rederive results for transition temperatures and order parameters
that are less transparent in previous replica calculations, 
or only arrived at by analogy with the Bethe lattice. 
Indeed, 
results from bifurcation theory enabled us to calculate
the Hessians 
for bifurcations from the 
symmetric solution
(which may not be dominant
all the way to the bifurcation point, as
is seen in the $k=3,4..$ solutions)
in $k$ replica Ising or Potts models
for any $k$ and $P(J)$. 
The number of 
putative spin glass solutions 
bifurcating at $\beta_2$ appeared to behave as in the
the infinite range model. In essence the investigation of a
spin glass on a $z-$regular
random graph is reduced to looking at the solutions
to the equation
\begin{eqnarray}
\vec \phi &=& \ \int \;  P (J) \otimes^k K \; dJ 
 \; \vec \phi^{z-1} \nonumber \\
          & lim \; k \rightarrow 0&.
\end{eqnarray}
Further work along these lines is clearly both desirable and possible
for the Ising and other Potts models, with the nature 
of replica symmetry breaking in this short range (but still mean
field) model being perhaps the most important question. We have made
no attempt here to follow the various solution 
branches that bifurcate at $\beta_2$,
for instance, which would shed light on the low temperature phase,
particularly if calculations for arbitrary $k$ were still possible.
An understanding of the role of the first order solutions that
appear for $k>2$ and 
their failure to influence the $k \rightarrow 0$ limit is
also still missing. We emphasize again that the 
transitions 
in the thin graph model appear to be identical
with those in the Ising replica magnet where it is the
$k=2$ transition that marks the boundary between
first order transitions and the behaviour seen in
the quenched model at $k=0$.
 
We investigated the models numerically in some detail.
Quantitatively, transition temperatures were in agreement
with those calculated for all the models. The extraction
of the critical exponents for the Ising model showed
clear mean field behaviour, and the $p$-independence
of the spin glass transition temperature was also apparent.
Qualitatively, the $P(q)$ measured in the various models
backed up the mean field picture of the phase transition
with a continuous distribution for the Ising model
and sharper features for $Q>3$.

In summary, spin glasses on thin graphs offer a promising arena
for the application of ideas from matrix models, large-$n$
calculations in field theory
and bifurcation theory. The tensor (or near-tensor) product
of the inverse propagator allows some quite general expressions
to be derived for the Hessian in the saddle point equations
and offers a powerful line of attack on questions such as 
replica symmetry breaking. As a subject for numerical simulations
they offer the great advantage of mean field results with no
infinite range interactions and no boundary problems. 

It is worth 
remarking in closing that one is not limited to 
the mean-field theory with the current methods. It is possible
to ``fatten'' the graphs analytically by increasing the size
of the matrices in the saddle point equations $N=2,3 \ldots$
as has already been done with some success for Ising models
coupled to two dimensional gravity \cite{hik}, 
which is equivalent to looking at
the models on an annealed ensemble of planar (fat)graphs.
This reduces the fractal dimension of the graphs from
infinity in the mean field case to more realistic values.
Numerical simulations of the ferromagnetic transition \cite{xx}
and possible spin glass transition \cite{0} have already been carried out
on a quenched ensemble of such planar graphs and in the spin glass case
$P(q)$ still presents a mean-field-like appearance. It would be very
interesting to say something analytically
about the nature of the low
temperature phase in such a non-mean-field spin glass.

\section{Acknowledgements}

The bulk of the simulations were carried out 
on the Front Range Consortium's
208-node Intel Paragon located at NOAA/FSL in Boulder.
Some simulations were also performed on
the Cray T3D at the Conrad Zuse Centrum in Berlin.
Some of the Binder's cumulant analysis was carried out using
programs written by A. Krzywicki.
CFB is supported by DOE under
contract DE-FG02-91ER40672, by NSF Grand Challenge Applications
Group Grant ASC-9217394 and by NASA HPCC Group Grant NAG5-2218.
PP is supported by EPSRC grant GR/J03466 and WJ thanks the
Deutsche Forschungsgemeinschaft for a Heisenberg fellowship.
CFB and DAJ were partially supported by NATO grant CRG910091.

\centerline{\bf Figure Captions}
\begin{description}
\item[Fig. 1] The crossover in Binder's cumulant
for the overlap in the Ising
spin glass calculated with individual
disorder averages. Only the $250,500$ and $1000$ lattice data
is plotted for clarity. The lines are drawn only to guide the eye
and are {\it not} the best fit curves used to determine
the crossing points.
\item[Fig. 2] The crossover in Binder's cumulant 
for the overlap calculated with an overall average.
Again, only the $250,500$ and $1000$ lattice data
is plotted and the lines are to guide the eye only.
\item[Fig. 3] The specific heat for the Ising spin glass.
\item[Fig. 4] The spin glass susceptibility $\chi_{sg}$.
\item[Fig. 5] The linear susceptibility $\chi_M$.
\item[Fig. 6] $P(q) \;  vs \;  q$ for $p=0,0.3,0.5$ on graphs of 500
vertices at $\beta =1.2$, deep in the spin glass phase.
\item[Fig. 7] $P(q) \;  vs \;  q$ for the same $p$ on graphs of 500
vertices at $\beta =0.9$, close to $\beta_{SG}$. The
histogram for $p=0$, which deviates slightly, is labelled. 
\item[Fig.8] $P(q) \;  vs \;  q$ for the three state Potts model at $\beta=2.2$.
Data from graphs with $500$ and $1000$ vertices are plotted.
\item[Fig.9] $P(q) \;  vs \;  q$ for the three state Potts model at $\beta=3.5$.
Data from graphs with $500$ and $1000$ vertices are plotted.
\item[Fig.10] $P(q) \;  vs \;  q$ for the four state Potts model at $\beta=4.4$.
Data from graphs with $500$ and $1000$ vertices are plotted.
\item[Fig.11] The specific heat for the $Q=4$ Potts spin glass. As the 
transition (determined from the Binder's cumulant for $q$)
lies above the specific heat peak the values for the
larger lattices are only measured on this side. 
\end{description}

\bigskip
\begin{thebibliography}{99}
\bibitem{0} C. Baillie, D. A. Johnston and J-P. Kownacki, Nucl. Phys. {\bf B432} (1994) 551. 
\bibitem{1} L. Viana and A. Bray, J. Phys. {\bf C 18} (1985) 3037.
\bibitem{1a} M. Mezard and G. Parisi, Europhys. Lett. {\bf 3} (1987) 1067;\\
            I. Kanter and H. Sompolinsky, Phys. Rev. Lett. {\bf 58} (1987) 164;\\ 
            K Wong and D. Sherrington, J. Phys. {\bf A20} (1987) L793;\\
            K Wong and D. Sherrington, J. Phys. {\bf A21} (1988) L459;\\
            C. de Dominicis and Y. Goldschmidt, J. Phys. {\bf A22}
             (1989) L775;\\
            C. de Dominicis and Y. Goldschmidt, Phys. Rev. {\bf B41} (1990) 2184;\\
            P-Y Lai and Y. Goldschmidt, J. Phys. {\bf A23} (1990) 399.
\bibitem{1b} N. Persky, I. Kanter and S. Solomon, ``Cluster Dynamics
             for Randomly Frustrated Systems with Finite Connectivity'',
             Racah Institute preprint, to appear in Phys. Rev. Lett.
\bibitem{3} H. A. Bethe, Proc. Roy. Soc. {\bf A 150} (1935) 552;\\
            C. Domb, Advan. Phys. {\bf 9} (1960) 145;\\
            T. P. Eggarter, Phys. Rev. {\bf B9} (1974) 2989;\\
            E. Muller-Hartmann and J. Zittartz, Phys. Rev. Lett. {\bf
            33} (1974) 893.
\bibitem{4} S. Inawashiro and S. Katsura, Physica {\bf 100A} (1980) 24;\\
            S. Katsura, S. Inawashiro and S. Fujiki, 
            Physica {\bf 99A} (1979) 193;\\
            S. Katsura, Physica {\bf 104A} (1980) 333, {\it ibid} {\bf 141A} (1987)556;\\
            S. Katsura and S. Fujiki, J. Phys. {\bf C12} (1979) 1087;\\
             D. Thouless, Phys. Rev. Lett. {\bf 56} (1986) 1082;\\
             J. Chayes, L. Chayes, P Sethna and D. Thouless, Comm. Math
             Phys. {\bf 106} (1986) 41;\\
             P. Mottishaw, Europhys. Lett. {\bf 4} (1987) 333;\\
             K Wong and D. Sherrington, J. Phys. {\bf A20} (1987) L785.
\bibitem{4a} Y. Goldschmidt, Europhys. Lett. {\bf 6} (1988) 7.
\bibitem{4b} F. Peruggi, J. Phys. {\bf A16} (1983) L713.\\
             F. Peruggi, F. di Liberto and G. Monroy, J. Phys. {\bf A16}
             (1983) 811.
\bibitem{5} C. Bachas, C. de Calan and P. Petropoulos, J. Phys. {\bf A27} 
            (1994) 6121.
\bibitem{6} E. Brezin, C. Itzykson, G. Parisi and J.B. Zuber, 
            Commun. Math. Phys. {\bf 59} (1978) 35;\\
            M.L. Mehta, Commun. Math. Phys. {\bf 79} (1981) 327.
\bibitem{7} For a review see, J. Ambjorn, `` Quantization of Geometry'' 
            Les Houches 1994, hep-th/9411179.
\bibitem{8} J. Le Guillou and J. Zinn-Justin (editors), ``Large Order Behaviour
            of Perturbation Theory'', Amsterdam: North Holland  (1989).
\bibitem{9} B. Derrida, Phys. Rev. Lett {\bf 45} 79 (1980);\\
            Phys. Rev. {\bf B24} 2613 (1981).
\bibitem{9a} D. Sherrington, J. Phys. {\bf A13} 637 (1980);\\
             R. Penney, A. Coolen and D. Sherrington, J. Phys. {\bf A26}
             3681 (1993).
\bibitem{bif} M. Golubitsky and D. Shaeffer, ``Singularities and Groups
              in Bifurcation Theory'', Springer Verlag, New York (1985).
\bibitem{sk} S. Kirkpatrick and D. Sherrington, Phys. Rev. {\bf B17} 4384
             (1978);\\
             G. Parisi, Phys. Rev. Lett. {\bf 50} 1946 (1983).
\bibitem{10} M. Guo, R. Bhatt and D. Huse, Phys. Rev. Lett. {\bf 72} 4137
             (1994);\\
             H. Rieger and A. Young, Phys. Rev. Lett. 
            {\bf 72} 4141 (1994).
\bibitem{11} D. Gross, I. Kanter and H. Sompolinsky, 
Phys. Rev. Lett. {\bf 55} (1985) 304.
\bibitem{hik} E. Brezin and S. Hikami, Phys. Lett. {\bf B283} 203 (1992);\\        
              E. Brezin and S. Hikami, {\it ibid} {\bf B295} 209 (1992);\\
              S. Hikami, {\it ibid} {\bf 305} 327 (1993);\\
              S. Hikami, Physica {\bf A204} 290 (1994).
\bibitem{xx} D. Johnston, Phys. Lett. {\bf B277} (1992) 405;\\
            C. Baillie, K. Hawick and D. Johnston, 
Phys. Lett. {\bf B328} (1994) 251.
\end{thebibliography}
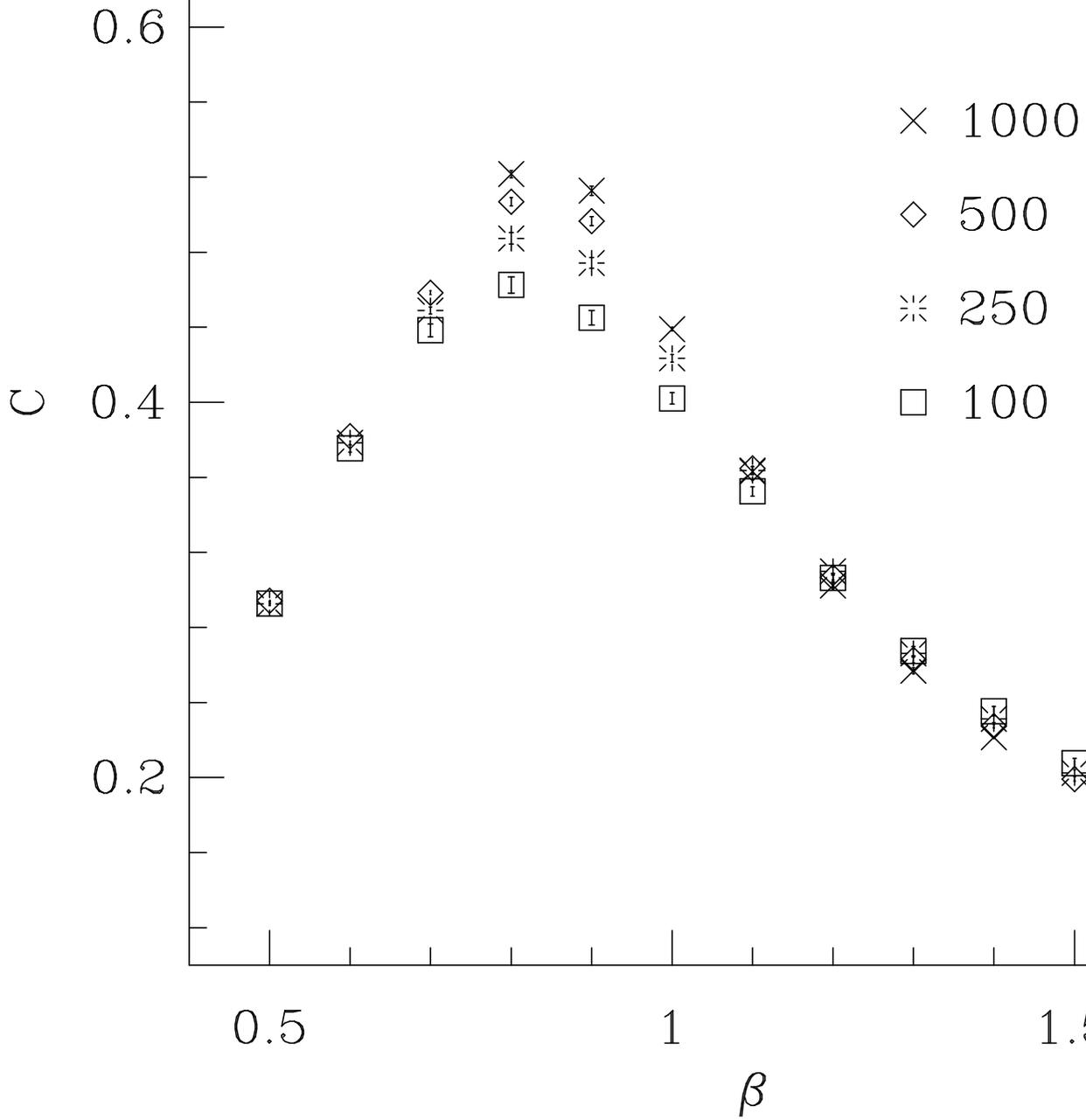
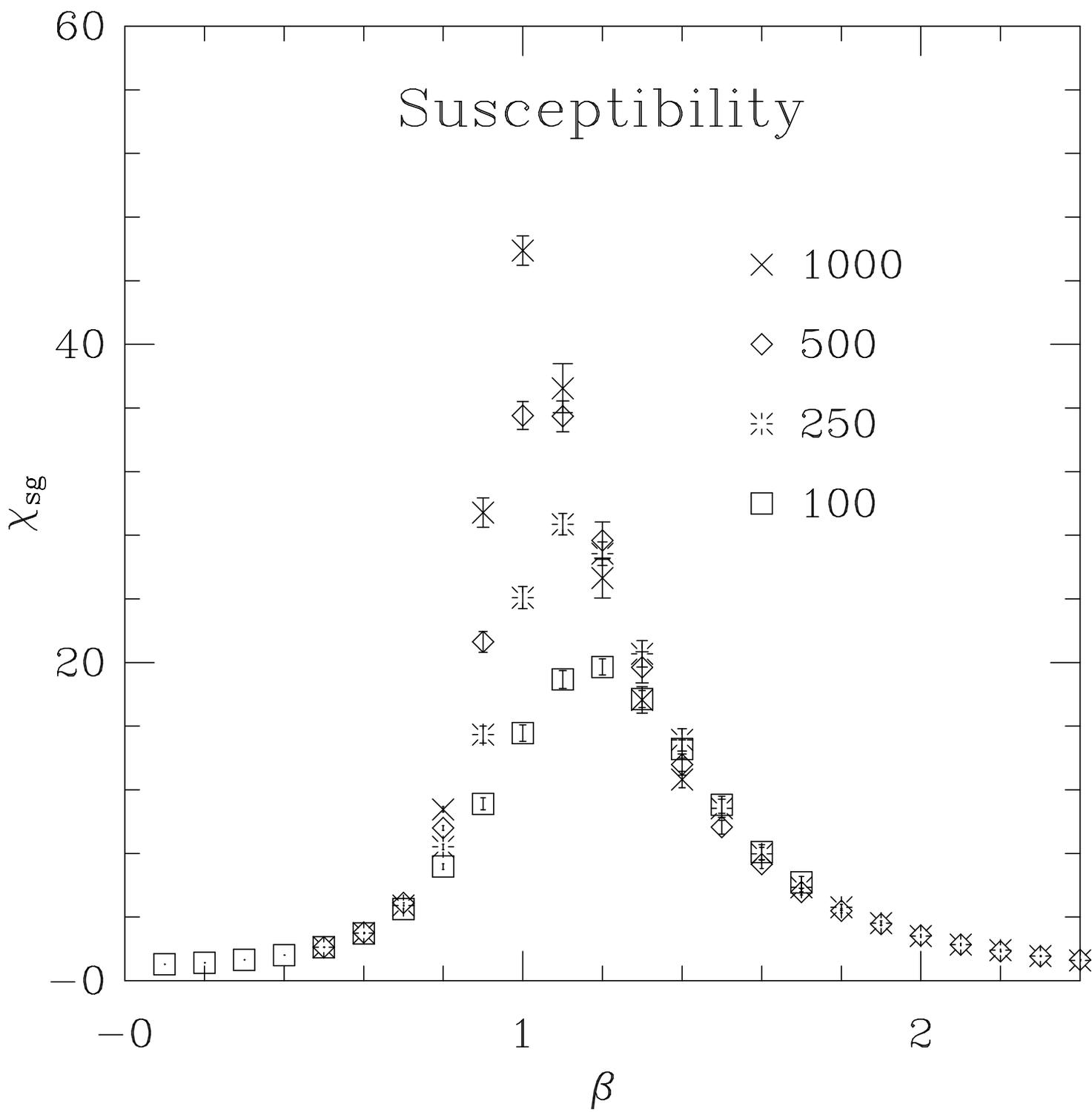
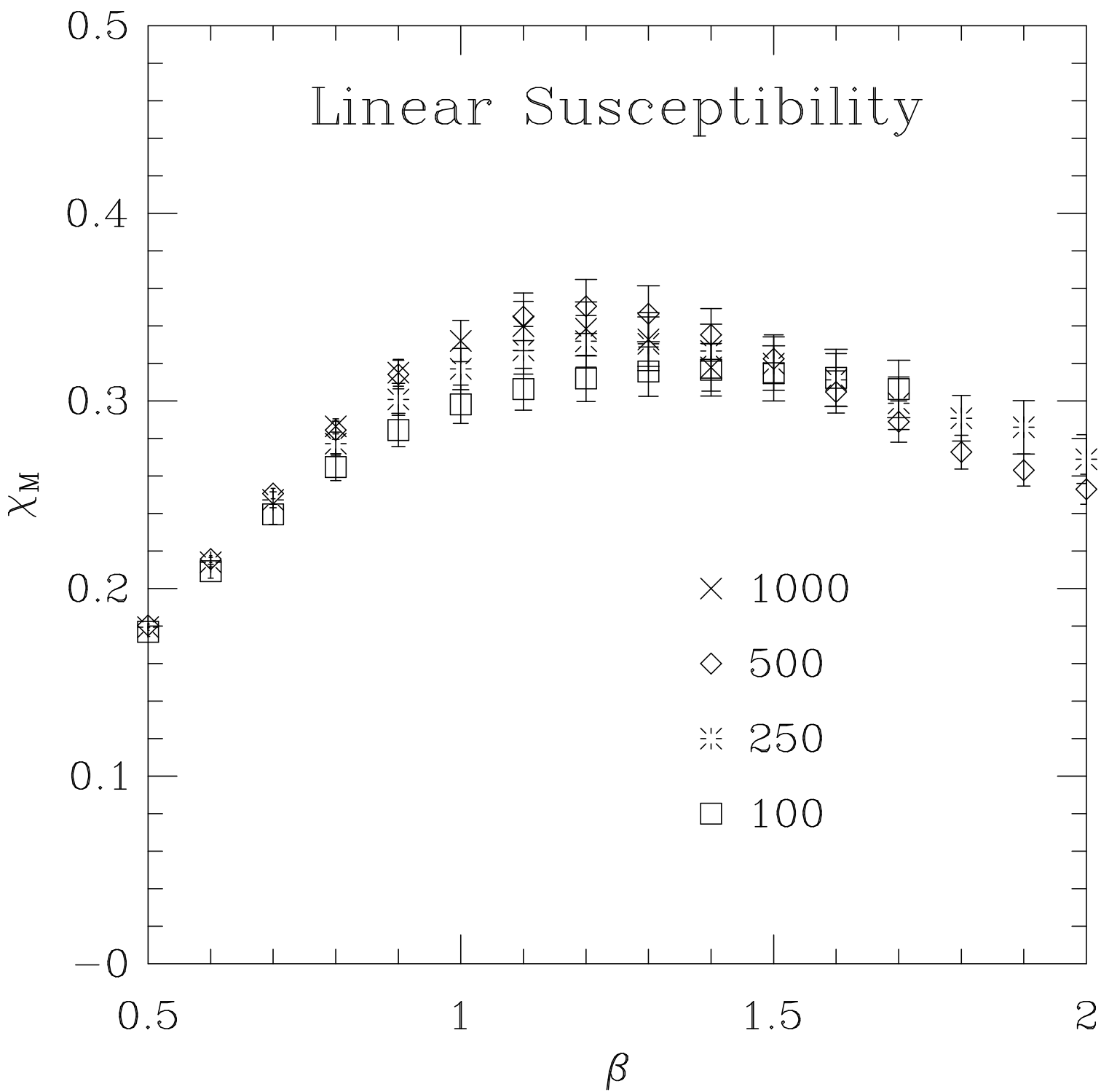
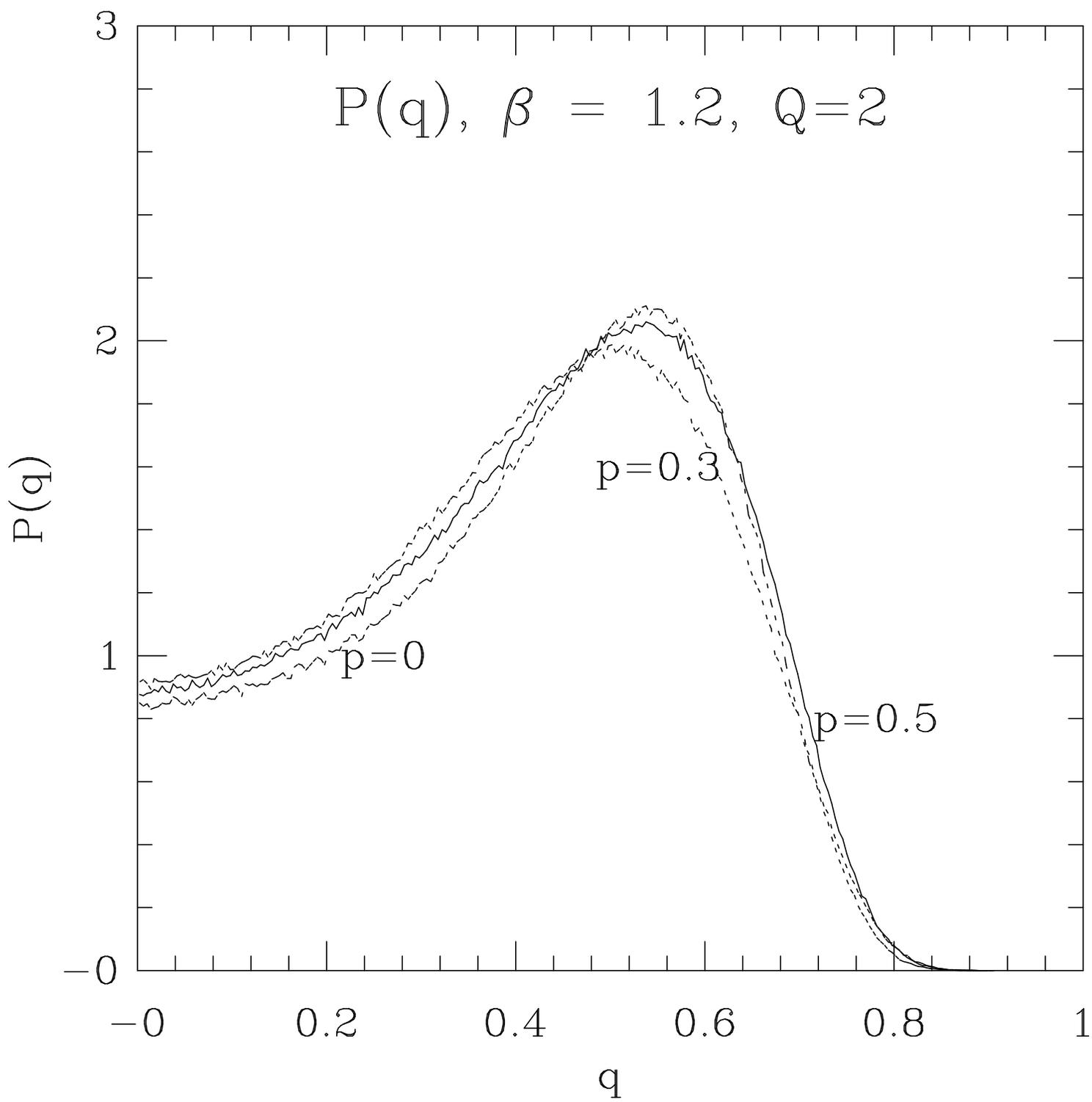
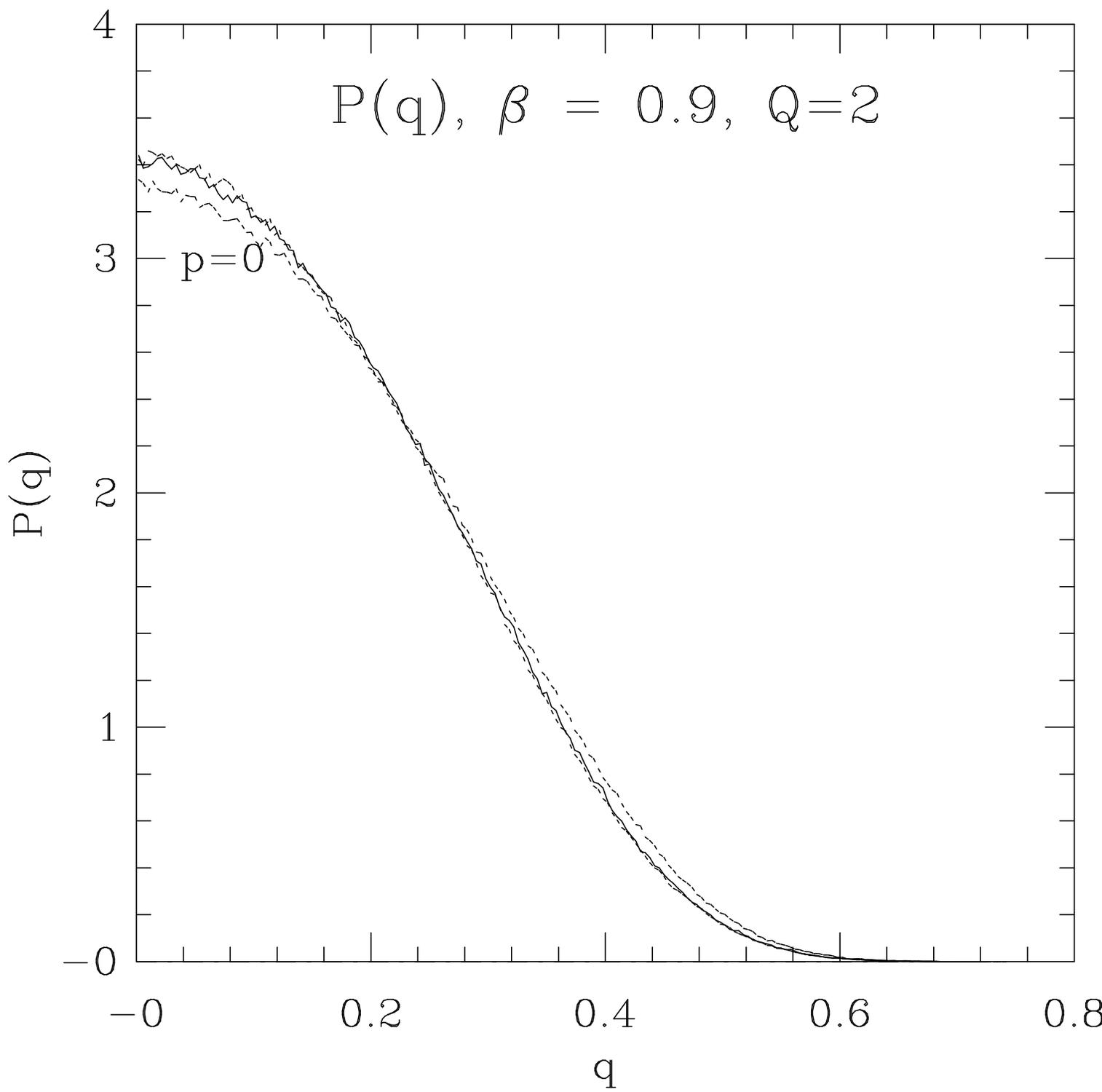
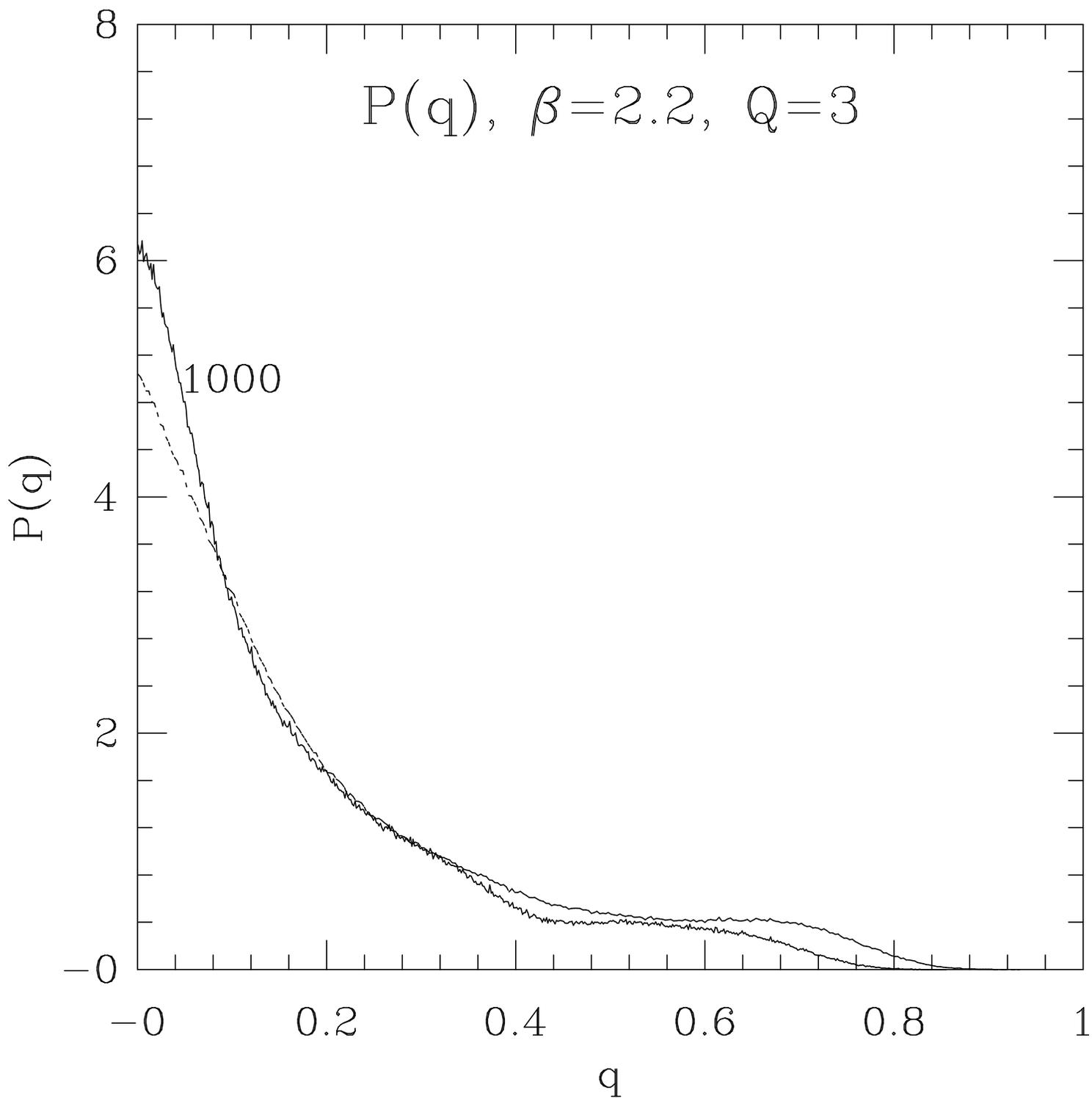
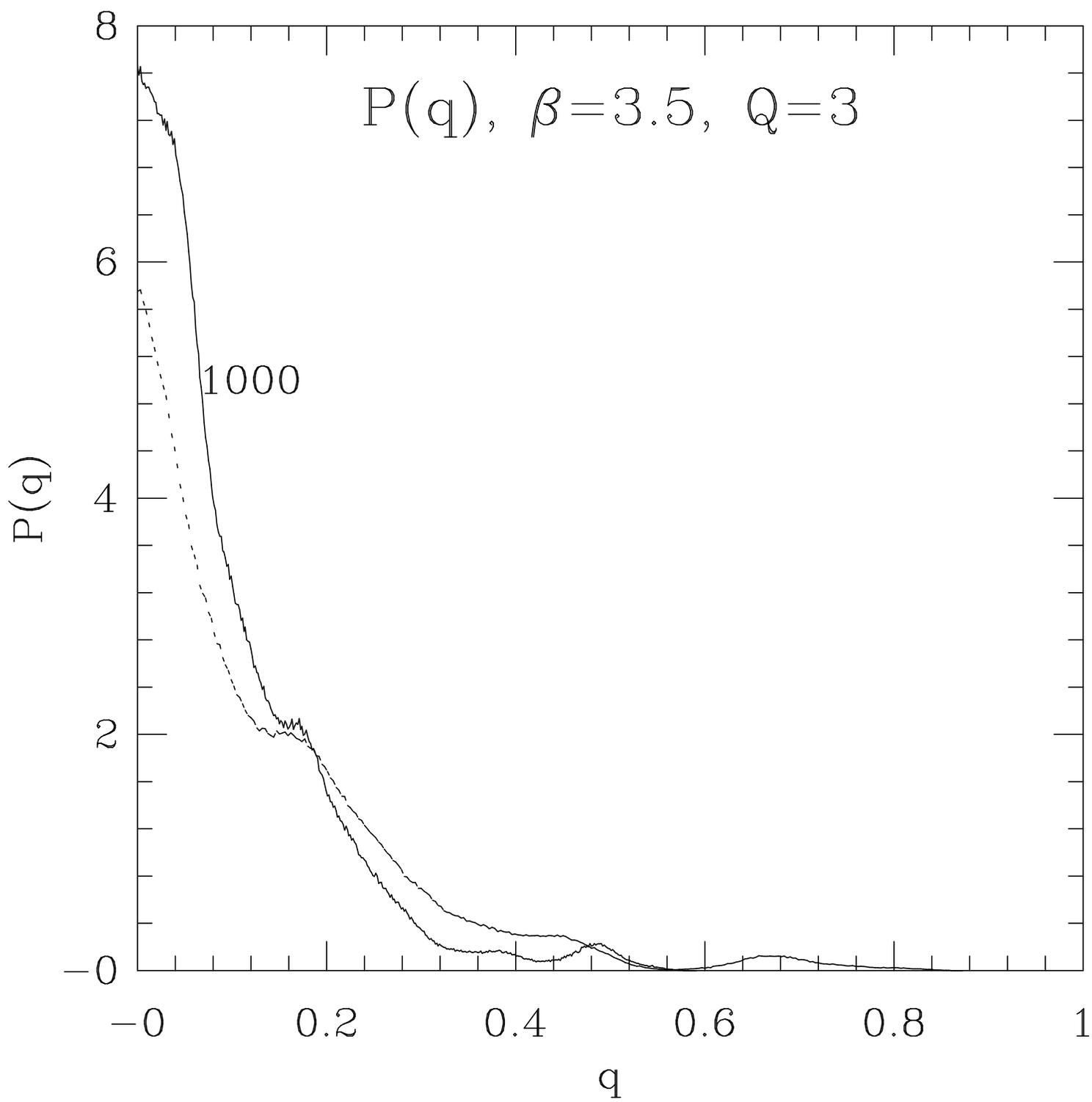
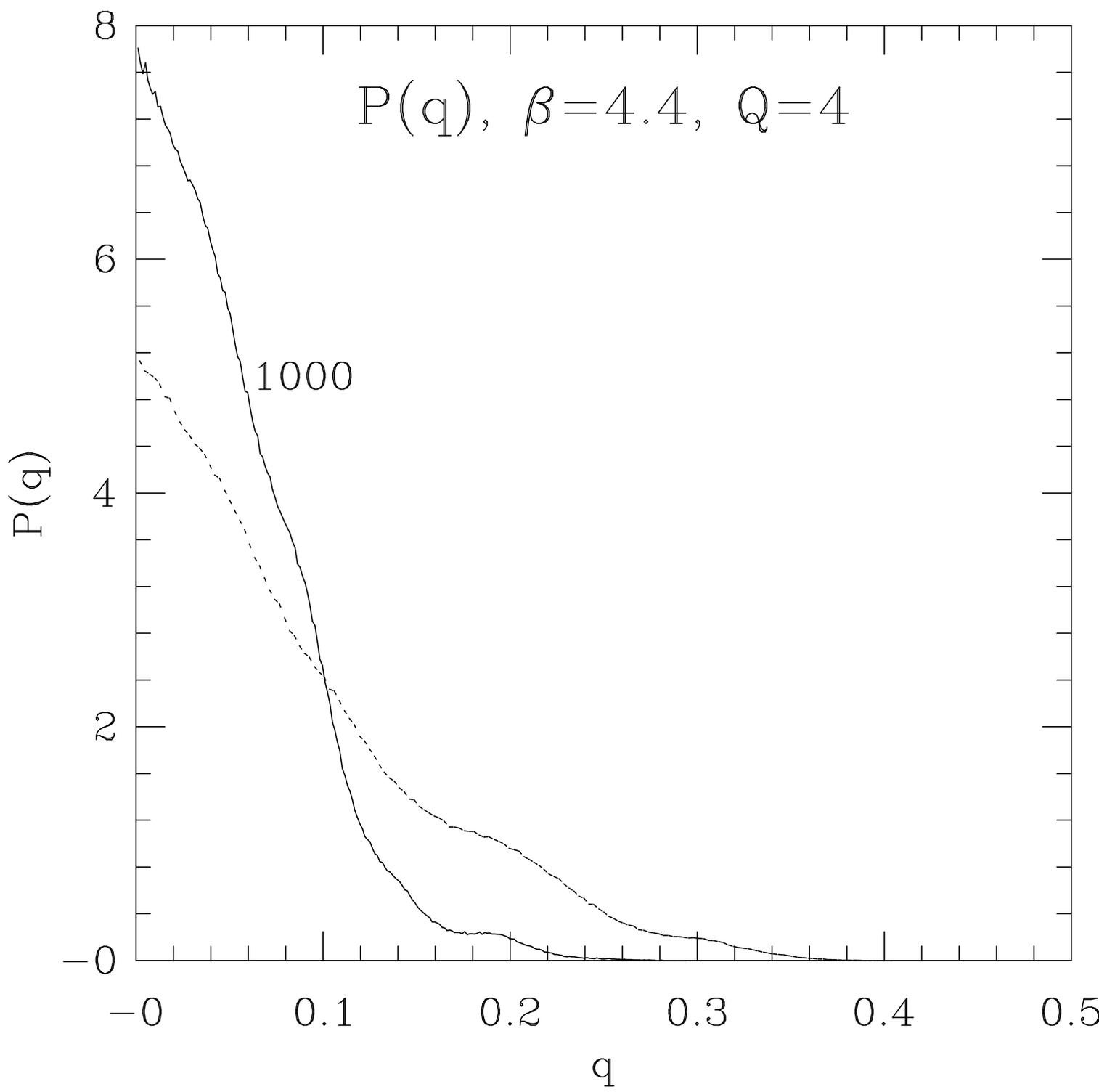
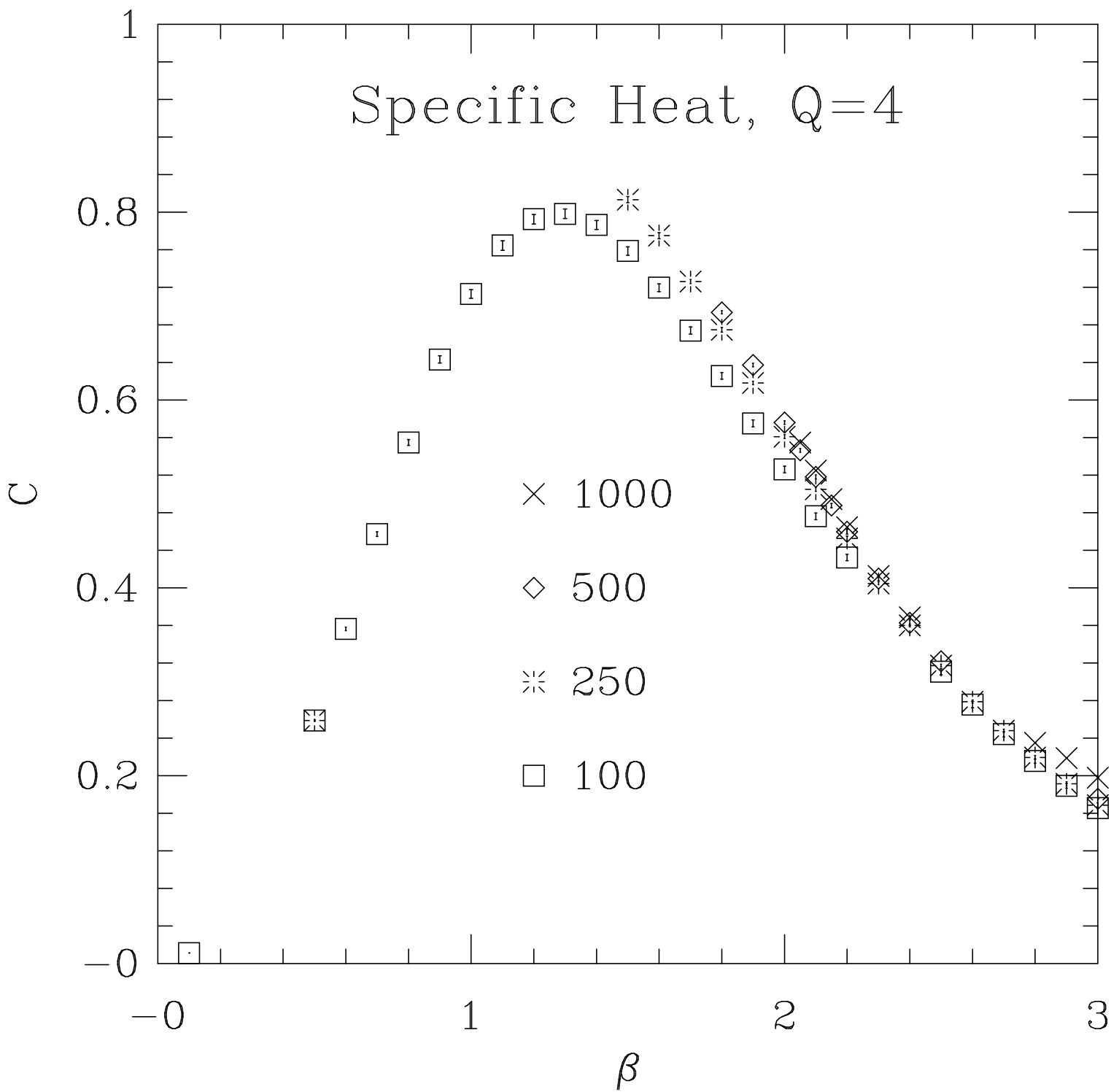
\end{document}